\documentclass[conference]{IEEEtran}
\IEEEoverridecommandlockouts

\usepackage{cite}
\usepackage{amsmath,amssymb,amsfonts}
\usepackage{algorithmic}
\usepackage{graphicx}
\usepackage{textcomp}
\usepackage{xcolor}
\usepackage[inline]{enumitem}
\usepackage{color,soul}
\usepackage{multirow}

\begin{document}

\title{Optimizing the Longhorn Cloud-native Software Defined Storage Engine for High Performance}

\author{
\IEEEauthorblockN{Konstantinos Kampadais\IEEEauthorrefmark{1}, Antony Chazapis, and Angelos Bilas\IEEEauthorrefmark{1}}
\IEEEauthorblockA{\textit{Institute of Computer Science} \\
\textit{Foundation for Reasearch and Technology - Hellas}\\
Heraklion, Crete\\
\{kampadais, chazapis, bilas\}@ics.forth.gr}
\thanks{\IEEEauthorrefmark{1}Also with Dept. of Computer Science, University of Crete.}
}

\maketitle

\begin{abstract}
Longhorn is an open-source, cloud-native software-defined storage (SDS) engine that delivers distributed block storage management in Kubernetes environments. This paper explores performance optimization techniques for Longhorn’s core component, the Longhorn engine, to overcome limitations in leveraging high-performance server hardware, such as solid-state NVMe disks and low-latency, high-bandwidth networking. By integrating ublk at the frontend, to expose the virtual block device to the operating system, restructuring the communication protocol, and employing DBS, our simplified, direct-to-disk storage scheme, the system achieves significant performance improvements with respect to the default I/O path. Our results contribute to enhancing Longhorn’s applicability in both cloud and on-premises setups, as well as provide insights for the broader SDS community.
\end{abstract}


\section{Introduction}

In cloud-native software architectures, the storage setup plays a significant role in determining the scalability and reliability properties of the whole system. Some core backend microservices, such as database engines, may handle this independently, by offering deployment options that distribute I/O operations across multiple instances, while automatically reacting to failures by exploiting internal data redundancies. However, all other microservices that require storage expect respective facilities to be provided by third-party software. For this reason, the cloud-native ecosystem is abundant with volume management solutions that either interface Linux-native or custom-built software defined storage (SDS) platforms to the APIs and deployment strategies of container orchestration environments. Implementing a cloud-native SDS has several benefits over choosing a similar product already available from a cloud provider, as it provides vendor independence, usually greater flexibility and cost efficiency, as well as advanced features that may not be part of standard offerings.

Longhorn \cite{longhorn} is a popular open-source, cloud-native volume manager, which implements its own distributed block storage system. It is a complete and independent SDS, handling internally all aspects related to capacity management, performance, fault tolerance, as well as interfacing with both Kubernetes and the end user. Longhorn is an actively developed and mature software, part of the CNCF software catalogue \cite{cncf}, however our installations have revealed that it currently lacks the ability to take advantage of the hardware performance available in servers that feature high-speed solid-state disks and high-bandwidth network connectivity.
This limits the applicability of Longhorn on setups with high I/O capabilities, as is often the case with the on-premise clouds or private systems colocated in a data center.

In this paper, we investigate a series of performance optimizations that collectively allow the system to achieve an order of magnitude better IOPS and bandwidth. We implement these optimizations in Longhorn's core, called the Longhorn engine \cite{longhorn-engine} (hereafter referenced simply as \textit{engine}), which is separately deployed in full for each managed volume. Each engine setup, consists of a \textit{controller} (data aggregator) and several \textit{replicas} (data storage endpoints). We remodel three points of the engine architecture, which we find to be the most critical to resulting performance:
\begin{enumerate*}[label=(\roman*)]
    \item the connectivity between the controller and the host operating system, where we use the \textit{ublk} framework \cite{ublk} instead of the current solution based on \textit{iSCSI/tgt} \cite{tgt},
    \item the communication between the controller and the replicas, where we employ a strategy that minimizes locks among involved threads, and
    \item the data storage scheme used by the replicas, which utilizes our custom direct-to-disk block management layer, named \textit{Direct Block Store} (DBS), instead of the default file-based implementation.
\end{enumerate*}

Our contributions in this work are:
\begin{enumerate*}[label=(\roman*)]
    \item we present how new operating system technologies, such as ublk, can be integrated into SDS designs,
    \item our implementation analysis offers a comprehensive methodology to diagnose performance bottlenecks and assess their impact in SDS stacks, and
    \item we introduce DBS, which may not use novel design concepts, but as an open source solution can serve as a useful utility to solve similar problems in related projects.
\end{enumerate*}
As we have practically affected all parts of the engine excluding the simplistic replication and data routing mechanisms, our modified system can facilitate future work by allowing for further performance analysis and optimizations.


\section{Related work}


Kubernetes provides an abstract API for specifying the storage requirements of services, through the use of \textit{PersistentVolumeClaim} objects. Such storage claims are monitored and served by storage plugins that implement the actual \textit{PersistentVolumes} which are then attached to running containers (in the Kubernetes nomenclature, the unit of execution is the \textit{Pod}, which may consist of one or more containers running in the same network namespace at the OS level). The details of the volume orchestration process in Kubernetes are defined in the Container Storage Interface (CSI) \cite{csi}. Actually, Kubernetes does not ship with a default CSI plugin as part of its installation, thus it is necessary for the administrator to add a compatible storage component for a fully working system.

Longhorn is one of several such CSI-compliant solutions, which has become favorable for its ease of deployment and use. Other popular CSI plugins include Rook \cite{rook}, which integrates with an SDS layer powered by Ceph \cite{ceph}, OpenEBS \cite{openebs}, which exposes the functionality of internal ``storage drivers'' that, in turn, implement distributed replicated storage, or can provision node-local storage from existing filesystem subdirectories, full or partitioned devices, or logical volumes via LVM or ZFS. All above systems are CNCF projects; the CNCF ecosystem also includes Piraeus \cite{piraeus}, a CSI-compliant interface to DRBD and Carina \cite{carina} that manages local RAID device groups. Beyond the CNCF, there are also numerous SDS commercial offerings compatible with Kubernetes, for which unfortunately limited technical background is publicly available, so they are outside the scope of this paper.


From a technical perspective, cloud-native volume managers can be categorized to broad groups, depending on
\begin{enumerate*}[label=(\roman*)]
    \item whether they implement their own SDS stack, or they provide an interface to an existing volume management software, and 
    \item whether they distribute data across nodes (usually managing redundancies, so to handle outages), or they confine their volumes on a single node.
\end{enumerate*}
Longhorn and OpenEBS fall in the first group of both criteria. 
An important part of these systems is how the SDS stack is exported to the OS. Longhorn, OpenEBS/cStor, and OpenEBS/Jiva incorporate a compatible iSCSI target at the top of the stack, and, in fact, share many similarities at the architectural level (iSCSI frontend, a controller acting as an I/O router, and distributed storage endpoints at different ndoes). iSCSI had been a common option to provide a virtual block device several years ago, however, newer technologies like NVMe-oF and ublk offer superior performance and efficiency, due to less overheads and kernel-bypass capabilities.

NVMe-oF is commonly used via SPDK (Storage Performance Development Kit) \cite{spdk} that provides userspace NVMe libraries and tools for efficient, low-latency storage operations. SPDK also includes bdev (block device), which uses a plugin architecture for implementing custom block drivers. On the other hand, ublk is a framework specifically made for creating block devices in userspace, which leverages the technology of io\_uring, a high-performance Linux kernel system call interface that uses ring buffers shared between the kernel and user space to submit and complete I/O requests efficiently, making it ideal for applications requiring scalable and fast I/O. Ublk and io\_uring, are offered in the latest Linux kernel (6.x) and included by default as part of most major Linux distributions (\textit{i.e.}, available in Ubuntu 24.10).

The NVME-oF solution has already been selected for the ``v2'' release of the Longhorn engine (currently in ``experimental'' status). OpenEBS follows a similar approach with the Mayastor driver \cite{mayastor} (currently under development). In both cases, NVME-oF---actually SPDK---requires a complete refactoring of the SDS code. In this work, we consider a solution based on ublk that is simpler and easier to integrate with an existing SDS. While ublk does not offer the same feature set (\textit{i.e.}, separating the block device point and the controller on different nodes and use RDMA for communication), experimental results suggest that the performance of the exported block device easily surpasses NVME-oF in the same setup \cite{longhorn_performance}.

\section{Longhorn architecture}

\begin{figure}[thb]
    \centering
    \includegraphics[width=0.90\linewidth]{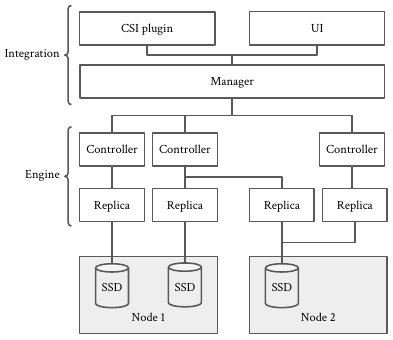}
    \caption{Longhorn components.}
    \label{fig:components}
\end{figure}

Longhorn components form a small web of microservices that provide distributed block storage for Kubernetes environments (Fig. \ref{fig:components}). At the core is the Longhorn engine, responsible for managing data replication and ensuring high availability across multiple replicas. Each Longhorn engine instance---a controller with associated replicas---implements a single volume. The Longhorn manager acts as the control plane, orchestrating the lifecycle of storage volumes, snapshots, and backups while interfacing with the Kubernetes API via the Longhorn CSI plugin. Additionally, the Longhorn UI provides a user-friendly dashboard for managing and monitoring storage operations. This work focuses on the engine service, which is the only component critical to performance since it implements the actual I/O path.

\begin{figure}[htb]
    \centering
    \includegraphics[width=0.57\linewidth]{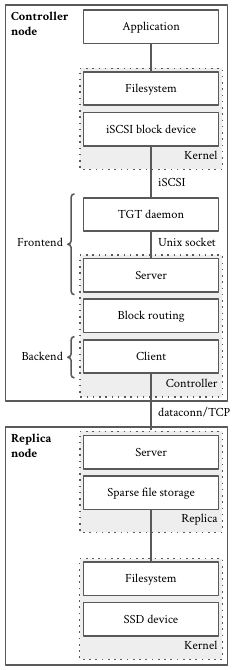}
    \caption{Original architecture of the Longhorn engine.}
    \label{fig:old-sys}
\end{figure}

Longhorn developers refer to the engine as ``world's smallest storage controller''. Indeed its design is simple and lightweight, consisting of three basic components (Fig. \ref{fig:old-sys}):
\begin{enumerate*}[label=(\roman*)]
    \item the \textit{frontend}, which is responsible for interfacing with the OS, so the controller's block API can be accessed seamlessly by applications over a virtual block device,
    \item the \textit{controller}, which routes I/Os to the replicas, functioning as a simple RAID controller (although only supporting mirroring), and
    \item the \textit{replica}, which stores the volumes' blocks in an actual device (the current implementation uses sparse files for storage).
\end{enumerate*}
In the distributed environment of Kubernetes, all these components are deployed in containers and communicate over the network. The frontend and controller are grouped together and run on the same node, while the replicas typically run on different nodes (one replica can be colocated with the frontend/controller).

The frontend creates a block device using the kernel iSCSI driver, by leveraging TGT, a third-party project that implements iSCSI targets in userspace. TGT has a plugin architecture that allows developers to materialize I/O operations using custom methods.
In order to connect TGT with the engine, the Longhorn TGT plugin uses a Unix socket and forwards each I/O request to the controller (the controller acts as the server, the TGT plugin as the client). Thus, each read and write issued in the exposed virtual iSCSI block device is redirected by the kernel to TGT, which, in turn, uses the Longhorn plugin to forward the request to the controller over a Unix socket.
TGT deployment is done by the controller. TGT is packaged along the controller, in the same container, and the controller includes a Golang wrapper that executes CLI commands on volume startup, which start TGT after the Unix socket from the controller's side is ready to accept connections.

The controller's basic responsibility is to accept the requests issued from the frontend, forward them to the replicas, and vice versa. The controller also employs a secondary out-of-band communication mechanism to receive management commands from the Longhorn manager. Such commands include volume start and stop, snapshot, and backup operations.

The layer of the controller that communicates with the replicas is internally called the \textit{backend}. Between the frontend and backend, the controller does not process requests (\textit{i.e.}, performs no erasure coding or deduplication). Each write is replicated to all replicas, and each read is served by one replica in round robin fashion. Note that each write creates multiple messages to replicas that all need to be executed before the command completes and the final response is sent back to the frontend.

The replica is the last componenent of the engine, responsible for storing the data in a physical medium. replicas consist of two basic layers: replica-to-controller communication and the storage mechanism. The communication part is implemented similarly to the controller's Unix socket server (albeit over TCP); the replica acts as the server, using multiple threads to serve TCP connections and handle received requests. The protocol used for exchanging commands and results is custom to Longhorn, internally called ``dataconn''.

The replica backing store receives read and write block requests that should be persisted to a physical medium. The default Longhorn implementation uses Linux sparse files for data storage, which effectively delegates space allocation to the filesystem and abstracts the underlying physical device, allowing it to write to diverse backends such as block devices, local storage, or cloud volumes.
Sparse files efficiently manage storage by only allocating space for blocks that have been written to. This reduces storage overhead for volumes with significant unused capacity, which is particularly useful in cloud environments where applications may allocate large volumes but only use a fraction of the space.

The engine supports snapshots, by creating a new sparse file for each snapshot that only records changes. The latest snapshot and ``version'' of each replica's storage is kept in a separate metadata file, in order to identify that replicas are consistent among each other. In the case of a faulty replica, the controller is responsible for identifying it and rebuilding it using data from the most up-to-date copy.

The engine is fully implemented in Golang. At the implementation level, the controller and replicas make heavy use of Golang channels as queues for assigning requests to connections and routing back responses. However, the code is very flexible and modular, allowing developers to easily change parts or build and integrate new functionality.

\section{Design and implementation}

\subsection{Methodology}

\begin{figure}[htb]
    \centering
    \includegraphics[width=0.77\linewidth]{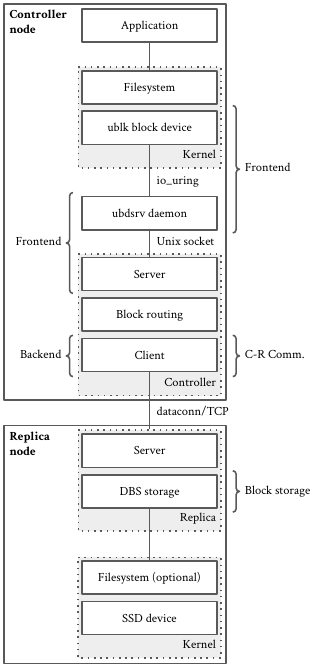}
    \caption{Modified architecture of the Longhorn engine. Changes described in the text are shown on the right.}
    \label{fig:new-sys}
\end{figure}

To work on the engine, we deploy it in a development setup of a single Linux server (Intel Core i7-10700 CPU, 16 GB RAM) with a local NVMe device (Samsung PM9A1). All components are built locally and run as native binaries (not in containers). Evaluation of different layers is done in a top-down approach by replacing I/O operations with no-ops at various places:
\begin{itemize}
    \item To measure the performance of the frontend, we replace the controller's backend read and write commands, so instead of routing to replicas, I/Os are immediately completed (\textit{null backend}).
    \item Once the frontend bottleneck is reduced, the impact of controller-replica communication is evaluated similarly, by modifying the replica's dataconn server to reply as soon as requests are received (\textit{null storage}).
\end{itemize}
For measuring IOPS and bandwidth, we use the \textit{fio} utility, configured to perform direct I/Os on the resulting block device. In our---fairly recent---development server, the full system running a single controller and a single replica reading and writing in the filesystem over an NVMe disk (no volume snapshots), cannot achieve a performance greater that 50k/25k read/write IOPS. In comparison, running fio directly on the filesystem yields about 400k read/write IOPS.
The elements of the modified design (Fig. \ref{fig:new-sys}) are described in the following sections.

\subsection{Frontend}

Using the null backend, we measured that the stand-alone frontend could not surpass 60k read/write IOPS, which is a number too low compared to modern I/O standards. This lead us to assume that the TGT-based implementation imposes a significant bottleneck. After analyzing the code and exploring the capabilities of this frontend option, we traced the issue largely to the fact that all communication is done synchronously.

While a different approach over iSCSI may have provided a viable solution, there are several other technologies available for exposing userspace block devices more efficiently and easier to integrate with the rest of the engine. We also considered ``legacy'' options such as NBD (Network Block Device) \cite{nbd} that minimizes the stack to only the essential components, as the kernel can directly connect to the controller. NBD can be configured to utilize multiple client-server threads, which also helped in raising overall performance. A test deployment with an NBD server in Golang incorporated directly in the controller, allowed us to reach over 100k IOPS at the frontend.

Next we considered NVMe-oF and ublk, both of which had already been discussed by Longhorn’s development team. NVME-oF will be part of Longhorn v2. The ublk path had been suggested as an option for the current version of the engine and a proof-of-concept (PoC) implementation already existed. We used this PoC as a foundation and tailored it to work with the latest version of Longhorn.

In the ublk framework, there are two basic components: ublk\_drv and ublksrv. The ublk\_drv is the Linux kernel module, responsible for IO command communication, copying of pages and various administrative tasks regarding the exposed block device, such as add, delete, and recover. Ublksrv is the userspace application that serves the I/Os via a driver module (similar to TGT). The Longhorn ublk PoC already included a compatible driver. Another powerful ublk feature is multiple frontend queues. This increases the queue-depth of incoming I/Os, providing significant performance gains. Multiple queues were not enabled in the original PoC, however our implementation includes it as a configurable option. Enabling the option allows the frontend to serve just over 500k IOPS in our development setup.

\subsection{Controller-replica communication}

Having alleviated the issue at the frontend, the next bottleneck was identified at the path that forwards the I/O requests to the underlying replicas. Sending the requests to the replica (null storage) drops the performance to approximately 100k read/write IOPS. This practically means that the communication mechanism (plus a negligible routing overhead) imposes a 4x performance penalty.

In the controller, each request issued on the frontend creates a thread that handles it. The implementation takes advantage of Golang channels and their thread-safe mechanisms. The controller manages dual TCP connections to each replica. Each connection uses two threads (send/recv) responsible for data transport, while a common thread (loop) implements a loop function that handles both incoming requests from the frontend and responses from the replicas.

\begin{figure*}[tb]
    \centering
    \includegraphics[width=0.92\linewidth]{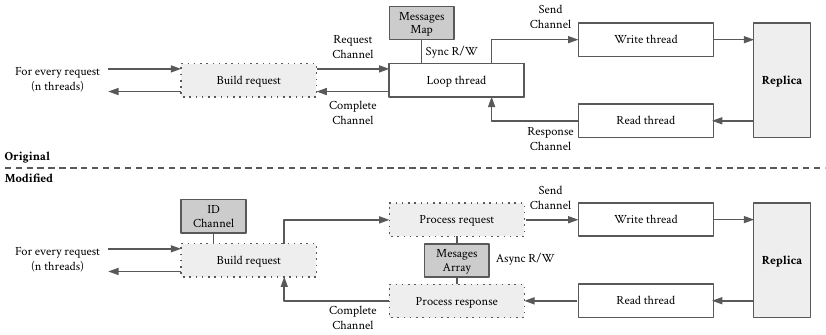}
    \caption{Original vs. modified controller-replica communication implementation.}
    \label{fig:crcfig}
\end{figure*}

After some experiments and code analysis, we found that this data path has limited potential. Although it is simple in design and uses most of the Golang features to its advantage, it fails to scale when multiple parallel requests are issued. The full data path is described as follows (Fig. \ref{fig:crcfig}):

\begin{enumerate}
\item Each request targeting a replica is translated into the appropriate form for controller-replica communication and inserted in the replica's Request Channel. The issuing thread sleeps until it is notified that the request is completed.
\item The loop function identifies that there is an issued request and reads it from the Request Channel, marks the message with a unique ID, and stores the message in a Golang map (Messages Map). The loop function forwards the message to the Send Channel. 
\item A send communication thread reads from the Send Channel the request and sends the data over the TCP connection.
\item From the replica's side a thread receives the data and serves the request. After the replica finishes serving, it sends the response back to the controller.
\item A receive communication thread receives the response and forwards the response to the Response Channel.
\item The loop function reads from the Response Channel and identifies the request that matches the received response using the ID as an index to the messages map. It changes any fields needed in the request and marks the request as completed, waking its corresponding thread.
\item The issuing thread finishes its execution, notifying the frontend about the result.
\end{enumerate}

We have traced the scalability problem to the use of a single thread running the loop function. This thread is responsible for handling all I/O operations, requests, and replies, creating a bottleneck. While its tasks are lightweight, the high volume of incoming requests and responses overwhelms the system, limiting performance to the capacity of a single thread. However, this implementation is necessary as the whole process is coordinated via a single Golang map (Messages Map). Maps are unable to do concurrent reads/writes, which is why requests and replies have to be processed sequentially (this also avoids locking to get the next available ID). In the default Longhorn implementation, the controller-replica communication performance is adequate because of the frontend bottleneck. Since our ublk-based frontend has significantly raised the number of I/Os that reach the controller, we need a new approach that avoids the single loop function.

Therefore, we have replaced the Messages Map with a simple fixed-size array (Messages Array) that holds a large predetermined number of IDs and a Golang integer channel. The Messages Array is sized equal to the maximum number of in-flight I/O operations we allow. The integer channel is initialized by populating it with the indexes of the Messages Array. These indexes act as unique request tokens. The modified data path now works as follows:

\begin{enumerate}
\item Each request targeting a replica is translated into the appropriate form for controller-replica communication.
\item The issuing thread acquires the next available ID from the Available IDs channel and stores the request's data in the Messages Array using the ID as an index. It then forwards the message to the Send Channel, and then sleeps until it is notified that the request is completed.
\item A send communication thread reads from the Send Channel the request and sends the data over the TCP connection.
\item From the replica's side a thread receives the data and serves the request. After the replica finishes serving, it sends the response back to the controller.
\item A receive communication thread receives the response and matches it to the corresponding request using the reply's ID as an index in the Messages Array. It changes any fields needed in the request and marks the request as completed, waking its corresponding thread.
\item The issuing thread finishes its execution, reinserts the request's ID into the Available IDs channel, and notifies the frontend about the result.
\end{enumerate}

The Golang channel guarantees that only one thread will acquire each unique ID. Since this ID is used as the index in the Messages Array, there are also no inconsistent read/write operations on the array, as each thread manipulates at most one index.

This approach eliminates the need for a loop function and scales up communication to handle much more incoming I/Os from the frontend. We also increased the number of concurrent connections to each replica from two to six; six was found to be the optimal number in our setup, effectively balancing system resource efficiency while maximizing the performance benefits of the new implementation. Now sending the requests to the replica (null storage) achieves almost double the IOPS (about 200k).

\subsection{Block storage}


Enabling the full path of I/Os to the device now reveals that another bottleneck is present at the replica's backing store, which is capped at 128k/38k read/write IOPS with all the refactoring done in the controller.
This is expected, as the default storage scheme has several shortcomings: Sparse files require a filesystem optimized for sparse file usage, as the latter has to maintain block allocation metadata and handle underlying fragmentation. Furthermore, each replica maintains a separate metadata file per volume with overall volume information, including the name and ``version'' of the latest data file. Management of such metadata in external files introduces overheads; we have verified that disabling write versioning raises the write IOPS significantly, almost to the level of reads. In addition, the performance degrades severely as the number of snapshots grows, as each snapshot is based on a new sparse file that only records new blocks in respect to the previous. Reads in volumes with many snapshots may have to go through the whole chain of sparse files in order to find the actual data block.

\begin{figure}[thb]
    \centering
    \includegraphics[width=0.64\linewidth]{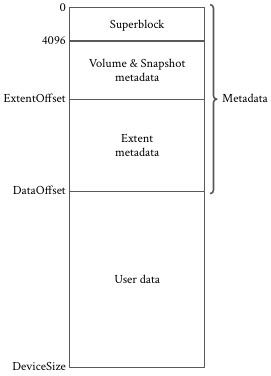}
    \caption{Internal structure of storage space managed by DBS.}
    \label{fig:dbs}
\end{figure}

To minimize the software layers involved in actual block storage and address the performance problem with multiple snapshots, we chose to redesign the block storage functionality, by introducing a custom, light-weight, direct-to-disk block storage implementation, offering a comprehensive framework for managing virtual volumes and snapshots on top of a physical block device or file. Each storage medium is managed by a low-level device layer and can contain multiple volumes, each having multiple snapshots. For interfacing, Direct Block Store (DBS) exposes an extensive API that provides high-level functionality for querying, managing, and manipulating volumes in a storage medium, as well as block-level I/O (read, write, unmap) for embedding into other applications. Additionally, DBS comes with a user-friendly command-line utility for device initialization, volume and snapshot management, and metadata queries. This allows performing operations such as adding, deleting, or renaming volumes, creating and cloning snapshots, and retrieving detailed information about the system's state outside Longhorn's context.

Logically, DBS divides the storage medium in fixed-size extents which correspond to the unit of allocation and management for each snapshot. Each 1 MB extent holds 32 4 KB blocks (which may or may not be all allocated) and belongs to a specific snapshot. Every volume is associated with a single snapshot; the latest in a volume-specific series of snapshots. A new volume always starts with a new snapshot; either empty or a clone of an existing one of any other volume (useful to restore data from a snapshot). Any volume can be deleted, which results in deleting the respective snapshot chain and deallocating all relevant extents. Any non top-level snapshot can be deleted; unique extents in that snapshot are merged with the next snapshot in the chain to maintain data integrity.

Internally, the storage medium is split into four regions (Fig. \ref{fig:dbs}): a superblock, a fixed-size region for volume and snapshot metadata, a variable-sized region for tracking the status of extents, and the remainder of the device, which stores actual user data. The snapshot extent maps (list of extents in order of position in snapshot) are not stored on the device, but are rather reconstructed at startup and kept in memory for maximum efficiency. Volume operations that only use in-memory extent metadata can proceed independently and are issued to the device in parallel to boost performance. Only writes to unallocated space require serialization, as they also update the superblock with the latest allocation mark. This also includes writes on previous snapshots extents, that are copied-on-write to new ones. DBS extensively utilizes bitmaps to quickly identify allocated regions for performance and---as in Longhorn's default storage scheme---uses direct I/O to bypass the OS cache when writing.

Overall, DBS is designed to deliver a stand-alone, high-performance block-level storage solution. It is written in Golang (using about 1000 lines of code) to be easily integrated into Longhorn, but may also prove suitable for other applications requiring direct access to raw storage with features that include volumes, as independent data domains, supporting point-in-time snapshots. In our development setup, DBS (deployed over the same filesystem) allows us to achieve end-to-end Longhorn performance of approximately 150k read/write IOPS, which corresponds to a 3x/6x improvement compared to the unmodified software.

\begin{table*}[tb]
\centering
\caption{IOPS Results}
\label{tab:iops}
\begin{tabular}{|l|r|r|r|r|r|r|r|r|} 
\hline
\multirow{2}{*}{}   & \multicolumn{2}{c|}{Upstream} & \multicolumn{2}{c|}{Ublk Frontend} & \multicolumn{2}{c|}{C-R Comm.} & \multicolumn{2}{c|}{DBS} \\
\cline{2-9}
                & Read & Write                      & Read & Write                      & Read & Write                      & Read & Write \\ 
\hline
Full engine     & \textbf{17k}  & \textbf{13k}      & 95k  & 27k                        & 110k & 27k                        & \textbf{112k} & \textbf{115k} \\ 
\hline
Without storage & 19k  & 19.5k                      & 100k & 100k                       & \textbf{129k} & \textbf{115k}     & \multicolumn{2}{l|}{$\rightarrow$} \\ 
\hline
Frontend only   & 20k  & 20k                        & \textbf{280k} & \textbf{255k}     & \multicolumn{4}{l|}{$\rightarrow$} \\
\hline
\end{tabular}
\end{table*}

\begin{table*}[tb]
\centering
\caption{Bandwidth Results (MB/s)}
\label{tab:bandwidth}
\begin{tabular}{|l|r|r|r|r|r|r|r|r|} 
\hline
\multirow{2}{*}{}   & \multicolumn{2}{c|}{Upstream} & \multicolumn{2}{c|}{Ublk Frontend} & \multicolumn{2}{c|}{C-R Comm.} & \multicolumn{2}{c|}{DBS} \\
\cline{2-9}
                & Read & Write                      & Read & Write                      & Read & Write                      & Read & Write \\ 
\hline
Full engine     & \textbf{300}  & \textbf{275}      & 1000 & 1000                       & 1250 & 1250                       & \textbf{1250}& \textbf{1250} \\ 
\hline
Without storage & 670  & 415                        & 1250 & 1250                       & \textbf{1250}  & \textbf{1250}    & \multicolumn{2}{l|}{$\rightarrow$} \\ 
\hline
Frontend only   & 750  & 415                        & \textbf{2000} & \textbf{2000}     & \multicolumn{4}{l|}{$\rightarrow$} \\
\hline
\end{tabular}
\end{table*}

\section{Evaluation}

\subsection{Setup}

To evaluate the system we deploy the controller and replica on separate nodes. In contrast to our development setup, this allows us to account for any effects related to the physical network. We use two identical nodes from our local cluster equipped with dual Intel Xeon E5-2620v2 CPUs, 128 GB RAM, connected via 10 Gbps Ethernet. Data is stored in a Samsung PM1733 NVMe drive. These machines have limited CPU core counts compared to state-of-the-art technology, however their specifications and performance is better aligned with typical VM offerings currently available in the cloud.

Actually, we initially evaluated our system in AWS, using two c5d.2xlarge EC2 instances, a cost-efficient option also used by Longhorn developers for publishing their benchmarks \cite{longhorn_report}.
However, EC2 instances have limited maximum provisioned IOPS, regardless of the hard drive used. As expected, the software performed to the machine's limit, reaching AWS's 40k IOPS cap. Overcoming these limitations in AWS requires using significantly more expensive EC2 instances. To harness the performance of the features described in this work, deployments must use instances and volumes at least 5 times more expensive than c5d.2xlarge. In any case, the behavior of the system on higher-performance cloud nodes is similar to the one we observe in our local setup.

\subsection{Results}

The results for IOPS and bandwidth are presented in Tables \ref{tab:iops} and \ref{tab:bandwidth} respectively. As in the previous section, in each experiment we do multiple runs to measure (shown as table rows): 
\begin{enumerate*}[label=(\roman*)]
    \item the \textit{full engine} performance, end-to-end, which includes writing blocks to the disk,
    \item the performance up to the replica \textit{without storage} using a null storage drive, where I/Os are immediately completed at the replica, and
    \item the \textit{frontend only} using a null backend, where I/Os are immediately completed at the controller.
\end{enumerate*}
We follow the same top-down approach, starting from upstream Longhorn and integrating each new feature in the same progression (shown as table columns). This indicates where the bottleneck is in each step and highlights how each solution (ublk frontend, controller-replica communication, DBS) contributes to the performance of the whole system.
For our experiments we use fio to measure IOPS (4k, random) and bandwidth (1 MB, sequential). All I/Os are direct to the virtual block device, bypassing kernel caches.

Starting with 17k/13k read/write IOPS when running the full stack, we isolate the first bottleneck at the frontend and measure it cannot achieve more than 20k IOPS using the upstream TGT-based solution. Integrating the ublk frontend yields a 28x boost, that allows the next layers of the engine to perform better, except for write IOPS, where the storage scheme of Longhorn fails to take advantage of the faster frontend. The bandwidth is also largely affected by the new frontend (almost 2.5x/4.8x for reads/writes compared to upstream), which enables the system to almost saturate the 10 Gbps links.
In general, these numbers support the general consensus among the community that ublk is an ideal framework to export SDS stacks to applications.

The next step is to evaluate the improved controller-replica communication implementation. Even with the ublk frontend, upstream Longhorn cannot achieve over 100k IOPS going up to the replica (null storage). Our modified communication scheme boosts performance by 29\%/15\% for reads/writes. There is still room for improvement, as we would ideally like to match the frontend performance, however it is enough for the engine to reach the full 10 Gbps bandwidth even with the default storage backend. The updated controller-replica communication also boosts random read IOPS to storage by 17\%. Write IOPS are not affected, which indicates that the bottleneck is in the storage backend itself (other runs have proved that this limitation is caused by write versioning). Lastly, integration of the DBS backend raises write IOPS to the level of reads; the whole modified system now performs an order of magnitude better than the default. Note that DBS is designed to keep the same performance level regardless of the number of volume snapshots.

\section{Conclusion and future work}

This paper demonstrates how we have overcome the limits of Longhorn’s engine in environments that feature high-speed storage and networking hardware, by making key modifications while maintaining compatibility with the existing system's architecture. By replacing the iSCSI-based frontend with a state-of-the-art ublk implementation, optimizing controller-replica communication, and introducing Direct Block Store (DBS) for efficient block storage at the replica layer, our version of the engine achieves up to an order-of-magnitude better IOPS performance in our evaluation setup.
Furthermore, the detailed analysis of each change's impact helps in understanding the accumulative effect the optimizations have in overall performance, which may be of interest to developers of similar systems; the technologies we use and the methodology we follow should be equally applicable to other SDS stacks.
All proposed enhancements have been submitted as pull requests to the upstream Longhorn repository, paving the way for integration into future releases and wider adoption by the cloud-native storage community.
Future work is planned to focus on identifying further areas of improvement in the controller, targeting both performance gains, as well as integration of advanced features at the level of replication and data routing.
DBS is also under active development; the roadmap includes inline compression and encryption to enhance the system’s utility for both cloud and on-premises deployments.

\section*{Acknowledgments}
The authors thankfully acknowledge the support of the European Commission under the Horizon Europe Programme through project DaFab (GA-101128693), as well as the European Commission and the Greek General Secretariat for Research and Innovation through project REBECCA (GA-101097224); REBECCA is managed by the Chips Joint Undertaking.
Credits for using Amazon Web Services were kindly provided by the National Infrastructures for Research and Technology (GRNET), under the Open Clouds for Research Environments (OCRE) Framework.

\bibliographystyle{IEEEtran}
\bibliography{main}

\end{document}